\documentclass[prd,twocolumn,showpacs,preprintnumbers]{revtex4}

\usepackage{amsmath}
\usepackage{epsfig}
\usepackage{dcolumn}
\usepackage{bm}

\newcommand{\tech}{t_{\scriptscriptstyle {\rm 1/2}}^{\scriptscriptstyle {\rm EC}}}
\newcommand{\cppn}{C_\nu(p_e,p_\nu)}

\begin{document}


\title{Low Energy Antineutrino Detection Using Neutrino Capture on
EC Decaying Nuclei}

\author{Alfredo G.~Cocco}
\email{alfredo.cocco@na.infn.it}
\author{Gianpiero Mangano}
\email{mangano@na.infn.it}
\affiliation{Istituto Nazionale di Fisica Nucleare - Sezione di Napoli \\
Complesso Universitario di Monte S.Angelo, I-80126 Napoli, Italy}

\author{Marcello Messina}
\email{marcello.messina@cern.ch}
\affiliation{Laboratorium f\"ur Hochenergiephysik - Universit\"at Bern\\
Sidlerstrasse 5,   CH-3012 Bern, Switzerland}

\date{\today}

\begin{abstract}
In this paper we present a study of the interaction of low energy
electron antineutrino on nuclei that undergo electron capture. We
show that the two corresponding crossed reactions have a sizeable
cross section and are both suitable for detection of low energy
antineutrino. However, only in case very specific conditions on
the $Q$-value of the decay are met or significant improvements on
the performances of ion storage rings are achieved, these
reactions could be exploited in the future to address the long
standing problem of a direct detection of Cosmological Neutrino
Background.
\end{abstract}


\maketitle

\section{Introduction}

In a previous paper~\cite{nostro} we have considered $\beta^\pm$
unstable nuclei as interesting candidates for very low energy
neutrino detection. Indeed, the crossed reaction with an incoming
(anti) neutrino has no energy threshold and the product of the
corresponding cross section times velocity goes to a finite
constant value at low neutrino velocities. Furthermore, the fact
that neutrinos have a non zero mass has important consequences on
the kinematics of the capture process and leads to the
possibility, at least in principle, to unambiguously detect the
very low energy Cosmological Neutrino Background
(C$\nu$B)~\cite{irvine}.

In this paper we address the closely related case of nuclei that
decay through the electron capture (EC) process, where the nucleus
of a neutral atom $A$ captures one bound electron and produces a
daughter atom $B$ and an electron neutrino
\begin{equation}
e^- + A^{+} \rightarrow B^* + \nu_e \rightarrow B + \nu_e + n\gamma \ .
\label{ecec}
\end{equation}
The atom $B$, initially in an excited state with a missing
electron in an inner atomic shell, decays electromagnetically and
releases a total energy $E_l$. By simple considerations it turns
out that $E_l$ is the captured electron binding energy in the
field of the daughter nucleus. The nucleus of atom $B$ can be
produced in an excited nuclear state as well. As we will see in
details in the following, the behavior of reaction~(\ref{ecec})
depends on the value of the mass difference between the parent and
daughter neutral atoms $Q_{EC}=M(A)-M(B)$, the value of $E_l$ and
the value of the neutrino mass.

Electron capture processes are suitable to detect electron
antineutrino via the two crossed reactions
\begin{equation}
\bar{\nu}_e + A \rightarrow B^- + e^+ \ , \label{ncbec}
\end{equation}
and
\begin{equation}
\bar{\nu}_e +  e^- + A^{+} \rightarrow B \ . \label{ncbecd}
\end{equation}

In the following we will analyze in details both these processes
and give an estimate of the antineutrino cross section that in
most cases does not require the evaluation of nuclear matrix
elements. We will also show under which circumstances the two
processes could be used to measure very low energy antineutrinos
and estimate the amount of the background due to competing
processes.

\section{Kinematics of antineutrino capture on EC decaying nuclei}
\label{section2}

 The behavior of reaction~(\ref{ncbec}) as a function of
the $Q$-value can be divided in three categories. In case $Q_{EC}
> 2m_e - m_\nu$ the neutrino capture process has no energy
threshold since the $Q$-value is large enough to allow the
creation of a positron in the final state even without the
contribution of the electron mass in the initial state of the EC
decay. On the other hand, if the $Q$-value satisfies the relation
$Q_{EC} > 2m_e + m_\nu$, the  $\beta^+$ decay becomes
energetically allowed. This case is described in details
in~\cite{nostro} and will not be treated here. There is therefore,
a range of values of $Q$ that is $2m_\nu$ wide and that would
allow the detection of antineutrino with an arbitrarily small
energy. Transitions falling in this category would have the
remarkable property of a unique signature, since the positron in
the final state of reaction~(\ref{ncbec}) can be used to tag the
antineutrino capture interaction with respect to the spontaneously
occurring reaction~(\ref{ecec}). Finally, in the case of
antineutrino captured on nuclei having $Q_{EC} < 2m_e - m_\nu$,
reaction~(\ref{ncbec}) has a threshold on the energy of the
incoming antineutrino given by
\begin{equation}
E_\nu^{\rm thr} = 2m_e - Q_{EC} \ , \label{ecthr}
\end{equation}
and the energy of the outgoing positron in this case reads $E_e =
E_\nu + Q_{EC} - m_e$. Though this threshold prevents in this case
the use of this reaction to detect low energy antineutrino,
nevertheless, atoms undergoing EC decays in which $Q_{EC} \simeq
2m_e$ could still be used for that purpose.

As far as reaction~(\ref{ncbecd}) is concerned, the $Q$-value
plays a crucial role as well. For $Q_{EC}-E_l \ge -m_\nu$
reaction~(\ref{ncbecd}) has no energy threshold and moreover, the
nucleus $A$ is stable since the corresponding EC decay become
energetically allowed only if $Q_{EC}-E_l \ge m_\nu$. Once again,
there is a range of $Q$-values that is $2m_\nu$ wide and in which
reaction~(\ref{ncbecd}) has no threshold on the energy of the
incoming antineutrino. The process is also background free since
the EC decay is energetically forbidden. Unfortunately,
reaction~(\ref{ncbecd}), as it is, is forbidden by the lack of a
suitable final state. Using the Fermi golden rule
($w=2\pi/h\left|{\cal M}\right|^2 \rho_f(E)$) one gets that the
cross section depends on the number of available final states per
unit energy $\rho_l(E_\nu)=\delta(E_\nu+(Q_{EC}-E_l))$, which in
case of an incoming antineutrino at rest has only one possible
solution, $Q_{EC}-E_l=-m_\nu$. Despite of this, we can still
envisage at least two cases where the process might be allowed:
\begin{itemize}
\item[i)] there exists an excited state $B'$ having energy
$M(B)+E_\nu+Q_{EC}-E_l$; in this case the background reaction (EC
decay through the same channel) would be forbidden due to energy
conservation even in the limit of $E_\nu \rightarrow m_\nu>0$
\item[ii)] the captured electron is ``off-mass shell'' with an
effective mass given by $m_{\rm eff}=m_e-Q_{EC}+E_l-E_\nu$; this
could happen for example in a metal when the nucleus captures an
electron in the valence band, being in this case $E_l$ the mean
binding energy of valence electrons.
\end{itemize}
In the case of $Q_{EC}-E_l<-m_\nu$ reaction~(\ref{ncbecd}) could
be still triggered by antineutrinos with energy greater than
$E_\nu^{\rm thr}=-Q_{EC}+E_l$. As an example, we recall the
process pointed out in~\cite{mikelyan}, where high energy reactor
antineutrino and the electron are captured by a stable nucleus $B$
and produces a $\beta$--unstable nucleus $A$.

\section{Antineutrino capture cross section}

The electron capture process has been studied in details
in~\cite{bambinek}. The corresponding rate is given by
\begin{equation}
\lambda_{EC} = \frac{G_\beta^2}{2\pi^3} \sum_x n_x C_x(q_\nu) f_x
\ , \label{lambdaec}
\end{equation}
where the sum is over all atomic shells from which an electron can
be captured, $n_x$ is the relative occupation number of that shell
and $C_x(q_\nu)$ is the nuclear shape factor relative to the given
transition. The index $x$ labels the orbital electron
wave-function via the variable $\kappa_x$ given by the spherical
waves decomposition, as described in~\cite{bambinek}. To give an
example, nuclear transitions with no change in angular momentum
can be coupled only by $\kappa_x = \pm 1$ wave-functions, namely
$K, L_1, L_2, M_1, M_2...$ shells. Finally, the function $f_x$ is
the analogous of the integrated Fermi function of the $\beta$
decay and is given by the following expression
\begin{equation}
f_x = \frac{\pi}{2} q_x^2 \beta_x^2 B_x \ , \label{acinque}
\end{equation}
where $q_x = (Q_{EC} - E_l)/m_e$ is the neutrino energy, $\beta_x$
is the Coulomb amplitude of the bound-state electron radial wave-function and
$B_x$ is the associated electron exchange and overlap correction.

In full analogy with the procedure used in~\cite{nostro}, the
antineutrino capture cross section for reaction~(\ref{ncbec}),
$\sigma_{(2)}$, (in the following the cross section index will
refer to the corresponding process, as introduced in Eq.s
(\ref{ncbec}) and (\ref{ncbecd})) can be written as
\begin{equation}
\sigma_{(2)} v = \frac{G_\beta^2}{\pi} p_e E_e F(Z,E_e) \cppn \ ,
\label{sigma1}
\end{equation}
where $\cppn$ is the shape factor of the antineutrino capture
interaction and should be evaluated using prescription given
in~\cite{behrens}, while $F(Z,E_e)$ is the Fermi function for the
outgoing positron. It is worth noticing here that $\cppn$ and
$C_x(q_\nu)$ contain the same nuclear form factors. Moreover,
$\cppn$ can be expressed as a weighted sum of the $C_x(q_\nu)$
provided the correct superposition of free-particle and bound
state electron wave-functions are evaluated for each specific
case. We will show that in most of the cases $\cppn$ can be
obtained to a very good approximation using leading order
$C_x(q_\nu)$ terms.

Using Eq. (\ref{lambdaec}), it is possible to rewrite the cross
section (\ref{sigma1}) in the following way
\begin{equation}
\sigma_{(2)}\, v = 2\pi^2 \ln2\  p_e E_e
\frac{F(Z,E_e) \cppn}{\tech \ \sum_x n_x C_x(q_\nu)
f_x} \ . \label{ecsigma}
\end{equation}
According to the procedure used in~\cite{nostro} we define
a shape factor ratio ${\cal A}$ as
\begin{equation}
{\cal A} =  \frac{\sum_x n_x C_x(q_\nu) f_x}{p_e E_e F(Z,E_e)
\cppn} \ , \label{aec}
\end{equation}
where $q_\nu$ is the energy of the outgoing neutrino in the EC
decay (reaction (\ref{ecec})) and the subscript ($e$) refers to
the positron in the final state of reaction (\ref{ncbec}). The
antineutrino cross section can then be written as
\begin{equation}
\sigma_{(2)}\, v  = \frac{2\pi^2
\ln2}{{\cal A} \cdot \tech } \ . \label{sigma}
\end{equation}

In case of reaction~(\ref{ncbecd}) it is easy to show that the
antineutrino cross section, $\sigma_{(3)}$ up to a numerical
factor which depends upon initial and final angular momentum
multiplicity of the considered process, is given by
\begin{equation}
\sigma_{(3)}\, v = \frac{G_\beta^2}{\pi}
\sum_x n_x C_x(p_\nu) g_x \rho_x(E_\nu)   \ , \label{tone}
\end{equation}
where $\rho_x(E_\nu)$ is the number of available final states per
unit energy for an electron captured on the shell $x$ and
\begin{equation}
g_x\ = \frac{\pi}{2} \beta_x^2 B_x \ , \label{gx}
\end{equation}
is the analogous of~(\ref{acinque}). As both energies in the
initial and final states are given, $\rho_x(E_\nu)$ is a Dirac
delta function $\delta(E_\nu+Q_{EC}-E_{l(x)})$. Therefore,
incoming neutrinos are captured only if their energy is compatible
with the mass difference between initial and final states.

For reaction~(\ref{ncbecd}) the shape factor ratio is given by
\begin{equation}
{\cal A'} =  \frac{\sum_x n_x C_x(q_\nu) f_x}{\sum_x n_x
C_x(p_\nu) g_x \rho_x(E_\nu)} \ , \label{aecd}
\end{equation}
where the variable $q_\nu$ refers to the neutrino energy in the
final state of the EC process while $p_\nu (E_\nu)$ is the
momentum (energy) of the incoming neutrino in
reaction~(\ref{ncbecd}). Also in this case the cross section can
be written according to Eq.~(\ref{sigma}).

We will now show that the shape factor ratios ${\cal A}$ and
${\cal A'}$ can be evaluated to a very good approximation for
allowed decays and in an exact way for superallowed and forbidden
unique decays.

\subsection{Superallowed transitions}

In case of superallowed transitions the shape factor involved in
the neutrino capture process is given by
\[
C_{EC}(p_e, p_\nu) =  \left| ^AF^{\scriptscriptstyle (0)}_{\scriptscriptstyle 101} \right|^2\ .
\]
On the other hand, the electron capture proceeds only via capture
from $K, L_1, L_2, M_1, M_2...$ shells, being contributions
involving electron orbital momentum forbidden. This means that
\[
C_x(q_\nu) = \left| ^AF^{\scriptscriptstyle (0)}_{\scriptscriptstyle 101} \right|^2 \ \ \ \ \kappa_x = \pm 1
\]
and that the shape factor ratios can be easily written as
\begin{equation}
{\cal A} =  \frac{\sum_x n_x f_x}{p_e E_e F(Z,E_e)}\ , \ \
\ \ {\cal A'} = \frac{\sum_x n_x f_x}{\sum_x n_x g_x
\rho_x(E_\nu)} \, \label{asuper}
\end{equation}
where both expressions do not depend anymore on nuclear matrix
elements evaluation. We notice here that ${\cal A'}$ can be seen
as the squared neutrino energy mean weighted with the electron capture
probability of each shell. This is of course strictly true only in
the limit of incoming neutrino having an energy greater than the
K-capture threshold.

\subsection{Allowed transitions}

Using the same arguments of~\cite{nostro} it is easy to show that
in case of allowed transitions and neglecting the (small)
contribution of nuclear transitions with a large angular momentum
transfer, the electronic capture shape factors reduce to a single
term, hereafter denoted by $C_0$, that describes the lowest order
transition and is independent of the outgoing neutrino energy. We
have
\begin{equation}
\sum_x C_x n_x f_x \simeq C_0 \sum_x n_x f_x \ \ \ \ \ C_0 \simeq
C_{EC} \ .
\end{equation}
Up to a very good approximation, the antineutrino capture shape
factor ratio is given in this case by (\ref{asuper}).

\subsection{Unique K-th forbidden transitions}

\begin{figure}
\centering \epsfig{figure=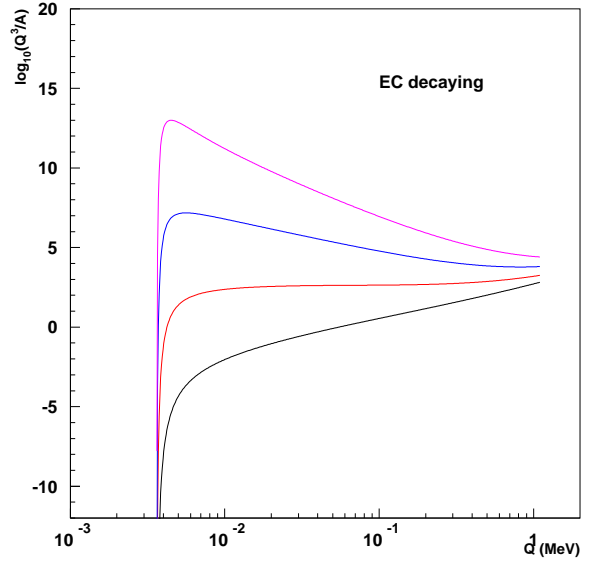,width=0.98\linewidth}
\caption{\label{aval} Values of $Q_{EC}^3/{\cal A}$ for EC
decaying nuclei at $E_\nu=2m_e$. The four curves represent from
bottom to top superallowed, first unique forbidden, second unique
forbidden and third unique forbidden transitions, respectively.
Curves are shown for $Z=20$ and the sharp cutoff at 3.6 keV is due
to the electron binding energy $E_l$ of the K shell electron
capture.}
\end{figure}

In case of K-th unique forbidden transitions and taking again only
dominant terms
\begin{equation}
C_x = \left| ^AF^{\scriptscriptstyle (0)}_{\scriptscriptstyle
LL-11} \right|^2 {\cal B}_L^{k_x} (p_x R)^{2(k_x-1)} (q_x
R)^{2(L-k_x)} \ ,
\end{equation}
where $L > k_x$ and numerical coefficients, here and in the following, can be
evaluated using prescriptions given in~\cite{bambinek}.
In case of capture from $s$-shells ($k_x = \pm 1$)
we obtain the simple form
\begin{equation}
C_x = \left| ^AF^{\scriptscriptstyle (0)}_{\scriptscriptstyle
LL-11} \right|^2 \frac{(q_x R)^{2(L-1)}}{(2L-1)!!} \ ,
\end{equation}
and the shape factor ratio can be written as
\begin{equation}
\frac{C_x}{C_\nu} = \frac{((2L-1)!!)^{-1} (q_xR)^{2(L-1)}}
    {\sum_{n=1}^{L} {\cal B}_L^n \lambda_n (p_e R)^{2(n-1)} (p_\nu
    R)^{2(L-n)}}\ ,
\label{atredici}
\end{equation}
again with no dependence on the nuclear form factors. The ratio
$\cal A'$ is again given by~(\ref{asuper}). In Fig.~\ref{aval} we
show the value of the ratio $Q_{EC}^3/{\cal A}$ in case of
superallowed and unique forbidden EC transitions for a specific
case.

\section{Estimating antineutrino capture cross section}

The cross section for reaction~(\ref{ncbec}) can be evaluated
using ~(\ref{sigma}) and following the procedure illustrated in
Section \ref{section2}. Numerical values for the constants
appearing in Eq.~(\ref{acinque}) and~(\ref{atredici}) can be found
in~\cite{bambinek}, while a description of the algorithm used to
compute the Fermi function is given in~\cite{nostro}. We report in
Table~\ref{ectab} the value of $\sigma_{(2)}$ for nuclei having
the largest product of cross section times lifetime for a specific
value of the incoming neutrino energy.
\begin{table}
\begin{tabular}{lcccc}
\hline
\hline
Isotope & Decay                        & $E_\nu^{\rm thr}$   & Half-life & $\sigma_{(2)}$  \\
        & ($J_i \rightarrow J_f$)      & (keV)               &  (sec)    & ($10^{-41}$ cm$^2$) \\
\hline
$^{  7}$Be  & $\frac{3}{2}^-\rightarrow\frac{1}{2}^-$ & 637.80 & $4.40\times 10^7$ & $6.80\times 10^{-3}$ \\
$^{  7}$Be  & $\frac{3}{2}^-\rightarrow\frac{3}{2}^-$ & 160.18 & $5.13\times 10^6$ & $1.16\times 10^{-2}$ \\
$^{ 55}$Fe  & $\frac{3}{2}^-\rightarrow\frac{5}{2}^-$ & 790.62 & $8.64\times 10^7$ & $1.55\times 10^{-5}$ \\
$^{ 68}$Ge  & $0^+ \rightarrow 1^+$                   & 916.00 & $2.34\times 10^7$ & $1.39\times 10^{-4}$ \\
$^{178}$W   & $0^+ \rightarrow 1^+$                   & 930.70 & $1.87\times 10^6$ & $5.14\times 10^{-4}$ \\
\hline
$^{ 41}$Ca  & $\frac{7}{2}^-\rightarrow\frac{3}{2}^+$ & 600.61 & $3.22\times 10^{12}$ & $8.35\times 10^{-9}$ \\
$^{ 81}$Kr  & $\frac{7}{2}^+\rightarrow\frac{3}{2}^-$ & 741.30 & $7.23\times 10^{12}$ & $2.40\times 10^{-9}$ \\
$^{100}$Pd  & $0^+ \rightarrow 2^-$                   & 693.68 & $3.14\times 10^{5}$  & $4.17\times 10^{-4}$ \\
$^{123}$Te  & $\frac{1}{2}^+\rightarrow\frac{7}{2}^+$ & 970.70 & $1.89\times 10^{22}$ & $5.40\times 10^{-15}$ \\
\hline
\hline
\end{tabular}
\caption{\label{ectab} Pure EC decaying nuclei with the largest
$\sigma{(2)} \cdot \tech$ value for neutrino capture processes of
Eq.~(\protect{\ref{ncbec}}). Cross section is evaluated for
incoming antineutrino energy of 1 MeV above reaction threshold and
in case of K shell capture. Allowed transitions (top) and
forbidden unique (bottom) are shown.}
\end{table}

\begin{figure}
\centering \epsfig{figure=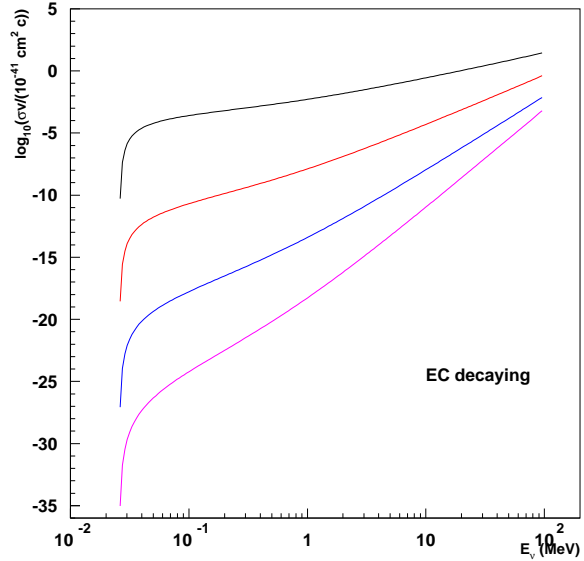,width=0.98\linewidth}
\caption{\label{figecxs} The product $\sigma v_\nu$ for EC decaying nuclei versus neutrino energy
for reaction~(\protect{\ref{ncbec}}). Typical values for
$\log(ft)$ values have been assumed~\protect{\cite{singh}} from top
to bottom as follows: allowed (5.5), first unique
forbidden (9.5), second unique forbidden (15.6) and third unique forbidden
(21.1). The curves refer to $Q_{EC}=1$ MeV, $Z=20$
and nuclear radius given by $R=1.2 A^{1/3}$ fm, where $A=2.5 Z$.}
\end{figure}

Antineutrino capture cross section behavior as a function of the
incoming antineutrino energy is shown in Fig.~\ref{figecxs} for a
specific case. As an example, we consider the case of $^7$Be,
which decays with a half-life of $\tech=53.22$ days~\cite{audi}
with $Q_{EC}=861.815 \pm 0.018$ keV~\cite{audi}. The difference in
the electron binding energy between $^7$Be and its daughter $^7$Li
is of $E_l = 54.8$ eV~\cite{firestone}, and can be neglected with
respect to the decay $Q$-value. The energy threshold, according
to~(\ref{ecthr}), is of $160.24$ keV. Assuming an incoming
antineutrino with energy of $100$ eV above the energy threshold we
have that
\begin{equation}
\sigma_{(2)} = 2.0 \cdot 10^{-48}\ {\rm
cm}^2 \ ,
\label{hypo}
\end{equation}
in case of the $3/2^- \rightarrow 3/2^-$ transition.

Similarly, the antineutrino cross section for
reaction~(\ref{ncbecd}) can be easily written using
~(\ref{asuper}). For an electron captured from the K--shell we can
write (we recall that $q_\nu$ is outgoing neutrino in the EC
process)
\begin{equation}
 \sigma_{(3)}\, v = \frac{2\pi^2 \ln2}{q_\nu^2\ \tech}
 \rho_{\scriptscriptstyle \rm K}(E_\nu)\ ,
\end{equation}
with $\tech$ the half--life of the nucleus EC decay.

\section{Antineutrino capture versus EC decay rate}

Using Eq.s~(\ref{aec}) and~(\ref{aecd}), the ratio between
antineutrino capture ~(\ref{ncbec}) and~(\ref{ncbecd}) and EC
decay rates can be written as
\begin{equation}
 \frac{\lambda_\nu}{\lambda_{EC}} = \frac{2\pi^2}{{\cal A}^{{\scriptscriptstyle(}'{\scriptscriptstyle)}}}
 n_{\bar{\nu}}\ ,
\end{equation}
where $n_{\bar{\nu}}$ is the antineutrino density at the nucleus.
As an example, in case of reaction~(\ref{ncbec}) and superallowed
transitions we have that
\begin{equation}
 \frac{\lambda_\nu}{\lambda_{EC}} = 4\pi \frac{n_{\bar{\nu}}}{\sum n_x \beta^2_x B_x}
 \frac{p_e E_e F(Z,E_e)}{q_{\nu}^2} \ ,
\end{equation}
while for~(\ref{ncbecd}) and considering only K-shell capture in
superallowed and K-th unique transitions
\begin{equation}
 \frac{\lambda_\nu}{\lambda_{EC}} = 2\pi^2\frac{n_{\bar{\nu}}
\rho_{\scriptscriptstyle \rm K}(E_\nu)}{q_{\nu}^2} \ .
\end{equation}

It is possible to get an order of magnitude estimate for the rate
ratio using a simple argument. For reaction~(\ref{ncbec}), the
corresponding EC decay rate is related to the electron density at
the nucleus and to the electron neutrino phase space
\begin{equation}
 \frac{\lambda_\nu}{\lambda_{EC}} \simeq \frac{n_{\bar{\nu}}}{\left| \psi_e(0) \right|^2}
   \frac{p_e E_e F(Z,E_e)}{q_\nu^2} \ ,
\end{equation}
where $\psi_e(\vec{x})$ is the captured electron wave-function and
$E_e = (E_{\bar{\nu}} + Q_{EC}) - m_e$ is the outgoing positron
energy in the antineutrino capture process. The expected
antineutrino capture cross section is therefore, given by
\begin{equation}
   \sigma_{(2)} \simeq \frac{\lambda_{EC}}{\left| \psi_e(0) \right|^2}
            \frac{p_e E_e F(Z,E_e)}{q_\nu^2} \ ,
\label{ecxs}
\end{equation}
which depends only upon the experimental decay rate $\lambda_{EC}$
and on the electron wave-function at the nucleus. To test this
result, we re-evaluate the antineutrino capture cross section on
$^7$Be and compare the result with what we found in the previous
section. Assuming antineutrino having an energy of 100 eV above
threshold for the $3/2^- \rightarrow 3/2^-$
reaction 
and using expression~(\ref{ecxs}) we obtain
$\sigma_{(2)} = 2.88 \cdot 10^{-48}\ {\rm
cm}^2$ (compare with Eq.~\ref{hypo})
where a plain single particle hydrogen-like wave-function has been used to describe
K-shell electrons in the $^7$Be atom.

\section{Cosmological Antineutrino Background detection}

Recently, the possibility to detect the C$\nu$B using beta
unstable nuclei has received a great
interest~\cite{nostro,vogel,blennow}. In particular, this is
strictly related to the now well established experimental evidence
for neutrino mass from oscillation experiments. Though a direct
measure of the neutrino mass scale is still missing, results from
Large Scale Structure power spectrum and Cosmic Microwave
Background suggest an upper limit for the sum of the three
eigenstate masses at the 0.5 - 1 eV level (a lower value is
obtained if data from Lyman $\alpha$ clouds are included in the
analysis), see e.g. \cite{lesgourgues} for a review. In case
neutrino masses saturate this bound, Cosmological Neutrino
Background could be really within the experimental reach in the
near future. For an overview on perspectives for direct
measurements of the C$\nu$B see e.g.~\cite{ringwald}.

The detection of very low energy antineutrino using
reaction~(\ref{ncbec}) is of course problematic due to the
presence of the energy threshold in expression~(\ref{ecthr}). A
possible solution could be using the C$\nu$B as a target for
accelerated nucle. In this case the threshold energy is provided
by the energy of the accelerated nucleus in the C$\nu$B comoving
frame. From simple kinematical considerations, the minimal value
of the $\gamma$ factor (for a non-relativistic C$\nu$B electron
neutrino) reads
\begin{equation}
 \gamma_{\rm min} = \frac{E_\nu^{\rm thr}}{m_\nu} \ .
\end{equation}

It is worthwhile noticing here that EC decaying atoms show the
remarkable property of having a $Q$-value that depends on the
ionization degree of the parent atom. This is simply due to the
fact that the difference between the electron binding energies of
the parent and the daughter atoms depends on the total number of
electrons in the atom before and after the decay. This result is
well known and has been used for example, to evaluate the age of
the Universe using the Os-Re transition in case of fully ionized
atoms~\cite{osmium}.
In order to evaluate the change of the $Q$-value as a function of
the decaying nucleus we recall here that nuclear masses are
related to the atomic ones by the relation
\begin{equation}
M_N(A,Z) = M_A(A,Z) - Z \times m_e + B_e(Z) \ ,
\end{equation}
where $B_e(Z)$ is the total binding energy of all removed
electrons. Values of this quantity can be found in~\cite{huang},
while an useful parametrization is reported in~\cite{lunney}
\begin{equation}
B_e(Z) = 14.4381\ Z^{2.39} + 1.55468 \cdot 10^{-6} \times Z^{5.35}
\ {\rm eV} \ .
\end{equation}
In case of fully ionized atoms the effective $Q$-value is given by
\begin{equation}
 Q_{\rm eff} = Q_{EC} - m_e + \left[ B_e(Z) - B_e(Z-1) \right] \ .
\end{equation}
A completely ionized EC decaying atom whose $Q$-value is smaller
than $2m_e$ represents the best low energy antineutrino detector
one could achieve, since there are no competing backgrounds. The
process of Eq.~(\ref{ncbec}) only occurs in presence of an
electron antineutrino having an energy greater than $E_\nu^{\rm
thr}$.

As the relic antineutrino capture rate per nucleus can be
expressed as (we notice that $\tech$ is the half-life in the
nucleus rest frame)
\begin{equation}
 \lambda_\nu =
\frac{n_{\bar{\nu}}\, 2\pi^2 \ln2}{{\cal A} \cdot \tech }\ ,
\end{equation} the total rate is obtained by multiplying this
expression by the total number of accelerated nuclei ${\cal N}$,
where realistic values for present storage rings are ${\cal N} =
10^{13}$ and $\gamma = 100$. Assuming a transition having a value
of $Q_{\rm eff}$ of the order of the electronvolt this would lead
to an interaction rate in case of allowed transitions of the order
of $\lambda_\nu \simeq 10^{-18}$ s$^{-1}$, too slow to be
effectively detected even in absence of background due to the EC
decay of the nucleus (in case of a fully ionized beam).

The C$\nu$B detection using reaction~(\ref{ncbecd}) appears even
more difficult since for neutrinos having very small energy the
number of final states per unit energy $\rho_x(E_\nu)$ is
basically unknown. The atom in the final state has to have an
excess energy $Q_{EC}-E_l+m_\nu$ and this can only happen if this
energy can be radiated out via electromagnetic or phonon emission,
if the decaying atom is bounded in a solid. Photons emission can
be due either to atomic electrons or to nuclear level transition;
in the first case the typical energy lies in the eV-keV region
and, being $E_l$ in the same energy range, this implies that only
nuclei with a very small $Q$-value could be suitable for this
detection. In the second case, there should exist a nuclear level
that matches the energy difference. Furthermore, to avoid the
possibility that spontaneous EC decay are also allowed, these
levels must be $m_\nu$ above the transition $Q$-value by the
fine--tuned value $m_\nu$. Notice that in this latter case (EC
decay forbidden) there is a priori no easy way to evaluate the
cross section for reaction~(\ref{ncbecd}).

\section{Conclusions}
In this paper we have considered the interaction of low energy
electron antineutrino on nuclei that undergo electron capture
spontaneously. Depending on the $Q$-value, crossed processes where
a neutrino is in the initial state, could be in principle
exploited to measure low energy incoming neutrino fluxes from
astrophysical or cosmological sources. Using a method already
applied to neutrino captures on beta decaying nuclei in
\cite{nostro} to relate the nuclear shape factors of crossed
reaction, we have computed the expected neutrino-nucleus cross
section versus neutrino energy. The results shows that these
processes seem quite difficult to be used as a way to measure the
C$\nu$B, whose detection might be more promisingly pursued in the
future using beta decaying nuclei for sufficiently massive
neutrinos. Nevertheless, EC decaying nuclei could be an
interesting perspective for higher energy neutrino fluxes, if very
specific conditions on the $Q$-value of the decay are met or
significant improvements on the performances of ion storage rings
are achieved.

\bibliography{paper4}    

\begin{thebibliography}{15}
\expandafter\ifx\csname natexlab\endcsname\relax\def\natexlab#1{#1}\fi
\expandafter\ifx\csname bibnamefont\endcsname\relax
  \def\bibnamefont#1{#1}\fi
\expandafter\ifx\csname bibfnamefont\endcsname\relax
  \def\bibfnamefont#1{#1}\fi
\expandafter\ifx\csname citenamefont\endcsname\relax
  \def\citenamefont#1{#1}\fi
\expandafter\ifx\csname url\endcsname\relax
  \def\url#1{\texttt{#1}}\fi
\expandafter\ifx\csname urlprefix\endcsname\relax\def\urlprefix{URL }\fi
\providecommand{\bibinfo}[2]{#2}
\providecommand{\eprint}[2][]{\url{#2}}

\bibitem[{\citenamefont{Cocco et~al.}(2007)}]{nostro}
\bibinfo{author}{\bibfnamefont{A.~G.} \bibnamefont{Cocco}}
  \bibnamefont{et~al.}, \bibinfo{journal}{Journal of Cosmology and
  Astroparticle Physics} \textbf{\bibinfo{volume}{06}}, \bibinfo{pages}{015}
  (\bibinfo{year}{2007}).

\bibitem[{\citenamefont{Irvine and Humphreys}(1983)}]{irvine}
\bibinfo{author}{\bibfnamefont{J.}~\bibnamefont{Irvine}} \bibnamefont{and}
  \bibinfo{author}{\bibfnamefont{R.}~\bibnamefont{Humphreys}},
  \bibinfo{journal}{Journal of Physics G} \textbf{\bibinfo{volume}{9}},
  \bibinfo{pages}{847} (\bibinfo{year}{1983}).

\bibitem[{\citenamefont{Mika\'elyan et~al.}(1968)}]{mikelyan}
\bibinfo{author}{\bibfnamefont{L.~A.} \bibnamefont{Mika\'elyan}}
  \bibnamefont{et~al.}, \bibinfo{journal}{Sov. Jour. Nucl. Phys.}
  \textbf{\bibinfo{volume}{6}}, \bibinfo{pages}{254} (\bibinfo{year}{1968}).

\bibitem[{\citenamefont{Bambynek et~al.}(1977)}]{bambinek}
\bibinfo{author}{\bibfnamefont{W.}~\bibnamefont{Bambynek}}
  \bibnamefont{et~al.}, \bibinfo{journal}{Rev. of Mod. Phys.}
  \textbf{\bibinfo{volume}{49}}, \bibinfo{pages}{77} (\bibinfo{year}{1977}).

\bibitem[{\citenamefont{Behrens and B\"uring}(1982)}]{behrens}
\bibinfo{author}{\bibfnamefont{H.}~\bibnamefont{Behrens}} \bibnamefont{and}
  \bibinfo{author}{\bibfnamefont{W.}~\bibnamefont{B\"uring}},
  \emph{\bibinfo{title}{Electron Radial Wave Functions and Nuclear Beta Decay}}
  (\bibinfo{publisher}{Clarendon, Oxford}, \bibinfo{year}{1982}).

\bibitem[{\citenamefont{Audi et~al.}(2003)}]{audi}
\bibinfo{author}{\bibfnamefont{G.}~\bibnamefont{Audi}} \bibnamefont{et~al.},
  \bibinfo{journal}{Nucl. Phys. A} \textbf{\bibinfo{volume}{729}},
  \bibinfo{pages}{337} (\bibinfo{year}{2003}).

\bibitem[{\citenamefont{Firestone et~al.}(1996)}]{firestone}
\bibinfo{author}{\bibfnamefont{R.~B.} \bibnamefont{Firestone}}
  \bibnamefont{et~al.}, \emph{\bibinfo{title}{Table of Isotopes, 8th Edition}}
  (\bibinfo{publisher}{John Wiley \& Sons Inc., New York},
  \bibinfo{year}{1996}).

\bibitem[{\citenamefont{Lazauskas et~al.}(2008)}]{vogel}
\bibinfo{author}{\bibfnamefont{R.}~\bibnamefont{Lazauskas}}
  \bibnamefont{et~al.}, \bibinfo{journal}{Journal of Physics G}
  \textbf{\bibinfo{volume}{35}}, \bibinfo{pages}{025001}
  (\bibinfo{year}{2008}).

\bibitem[{\citenamefont{Blennow}(2008)}]{blennow}
\bibinfo{author}{\bibfnamefont{M.}~\bibnamefont{Blennow}},
  \bibinfo{journal}{Physical Review D} \textbf{\bibinfo{volume}{77}},
  \bibinfo{pages}{113014} (\bibinfo{year}{2008}).

\bibitem[{\citenamefont{Lesgourgues and Pastor}(2006)}]{lesgourgues}
\bibinfo{author}{\bibfnamefont{J.}~\bibnamefont{Lesgourgues}} \bibnamefont{and}
  \bibinfo{author}{\bibfnamefont{S.}~\bibnamefont{Pastor}},
  \bibinfo{journal}{Physics Reports} \textbf{\bibinfo{volume}{429}},
  \bibinfo{pages}{307} (\bibinfo{year}{2006}).

\bibitem[{\citenamefont{Ringwald}()}]{ringwald}
\bibinfo{author}{\bibfnamefont{A.}~\bibnamefont{Ringwald}},
  \eprint{arXiv:0901.1529[astro-ph.CO]}.

\bibitem[{\citenamefont{Bosch et~al.}(1996)}]{osmium}
\bibinfo{author}{\bibfnamefont{F.}~\bibnamefont{Bosch}} \bibnamefont{et~al.},
  \bibinfo{journal}{Phys. Rev. Lett.} \textbf{\bibinfo{volume}{77}},
  \bibinfo{pages}{5190} (\bibinfo{year}{1996}).

\bibitem[{\citenamefont{Singh et~al.}(1998)}]{singh}
\bibinfo{author}{\bibfnamefont{B.}~\bibnamefont{Singh}} \bibnamefont{et~al.},
  \bibinfo{journal}{Nuclear Data Sheets} \textbf{\bibinfo{volume}{84}},
  \bibinfo{pages}{487} (\bibinfo{year}{1998}).

\bibitem[{\citenamefont{Huang et~al.}(1976)}]{huang}
\bibinfo{author}{\bibfnamefont{K.-N.} \bibnamefont{Huang}}
  \bibnamefont{et~al.}, \bibinfo{journal}{At. Data and Nucl. Data Tables}
  \textbf{\bibinfo{volume}{18}}, \bibinfo{pages}{243} (\bibinfo{year}{1976}).

\bibitem[{\citenamefont{Lunney et~al.}(2003)}]{lunney}
\bibinfo{author}{\bibfnamefont{D.}~\bibnamefont{Lunney}} \bibnamefont{et~al.},
  \bibinfo{journal}{Rev. Mod. Phys.} \textbf{\bibinfo{volume}{75}},
  \bibinfo{pages}{1021} (\bibinfo{year}{2003}).

\end{thebibliography}

\end{document}